# An International and Multidisciplinary Teaching Experience with Real Industrial Team Project Development

MARTIN MELLADO, Universitat Politècnica de València,
EDUARDO VENDRELL, Universitat Politècnica de València,
FILOMENA FERRUCCI, Università degli Studi di Salerno
ANDREA ABATE, Università degli Studi di Salerno
DETLEF ZÜHLKE, Technische Universität Kaiserslautern
BERNARD RIERA, Université de Reims Champagne-Ardenne

This paper will present the design, objectives, experiences, and results of an international cooperation project funded by the European Commission in the context of the Erasmus Intensive Programme (IP, for short) designed to improve students' curricula. An IP is a short programme of study (minimum 2 weeks) that brings together university students and staff from at least three countries in order to encourage efficient and multinational teaching of specialist topics, which might otherwise not be taught at all. This project lasted for 6 years, covering two different editions, each one with three year duration. The first edition, named SAVRO (Simulation and Virtual Reality in Robotics for Industrial Assembly Processes) was held in the period 2008-2010, with the participation of three Universities, namely the *Universitat Politècnica de València* (Spain), acting as IP coordinator, the *Technische Universität Kaiserslautern* (Germany), and the *Università degli Studi di Salerno* (Italy). The *Université de Reims Champagne-Ardenne* (France) participated as a new partner in the subsequent edition (2011-2013) of the IP, renamed as HUMAIN (Human-Machine Interaction). Both editions of the teaching project were characterized by the same objectives and organizational aspects, aiming to provide educational initiatives based on active teaching through collaborative works between international institutions, involving industrial partners too. Six different enterprises participated in this experience, coming mainly from the automotive sector (i.e. Fiat) and from the food and beverage sector (as the industrial bakery *Dulcesol* or the champagne producer *Nicolas Feuillatte*), selected by the local university partner.

The IP was hosted each year in a different country. English was always the official language. In the total of the six editions, the project involved almost 150 students (77 in SAVRO and 72 in HUMAIN) from 10 European countries. A basic survey was organized at the end of each year in order to check the satisfaction of the students and to obtain suggestions in order to improve the IP. Almost 100% of students considered the IP course as a good and positive experience, both in personal and academic aspects. The aim of the paper is to illustrate the best practices that characterized the organization of our experience as well as to present some general recommendations and suggestions on how to devise computing academic curricula.

Categories and Subject Descriptors: **K.3.2 [Computers and Education]**: Computer and Information Science Education

## 1. INTRODUCTION

### 1.1 Computing skills and curricula

Computers and information technology (IT) represent the central technologies of our age, the main engines behind economic and social development. They have been dramatically changing how we live and work becoming so pervasive and essential in our society. As a matter of fact, computer systems control national infrastructures and services (e.g. transport, education, and health), financial systems, industrial manufacturing and distribution, entertainment. The strategic and efficient use of IT has been taking an increasingly decisive role for the competitiveness and innovation capability of industry and countries as well as the quality of our life.

Due to their significant practical impacts, Computing disciplines are not only very exciting and stimulating but also critical for our society. Properly educating and training the future generations of computing practitioners and researchers is crucial and can provide growth and competitive advantages.

Nevertheless, ensuring academic programs learning results satisfy actual real-world needs creates several significant challenges for Computing education that the community tries to address in several ways. Joint task forces of International Associations (e.g., ACM, IEEE-Computer Society, AIS) have been regularly appointed in the last 40 years to establish and revise international curricular guidelines for programs in Computing education (Computer Science, Computer Engineering, Information Systems, Information Technology and Software Engineering) with the aim of keeping those curricula modern and relevant. More recently the European Commission has activated some initiatives meant to understand the EU current as well as anticipated e-skills gaps and identify the most appropriate actions to address them, including guidelines for Computing education. ("e-Skills for the 21st Century: Fostering Competitiveness, Growth and Jobs).

Moreover, several international meeting points (e.g. ICSE Joint Software Engineering Education and Training, SIGCSE Technical Symposium, ACM International Computing Education Research Conference) regularly hold to discuss experiences, research results, and best practices meant to address the challenges characterizing Computing education.

Those activities are also meant to identify essential learning outcomes of Computing curricula to prepare computing graduates for professional careers. It is widely recognized that it is necessary educate them so that they are able:

— to analyse a problem and identify the best solution to solve it;
—to adopt a disciplined and systematic engineering approach for a cost/effective development and maintenance of software/IT product;
—to identify the best process that, under the constrained resources (e.g., time, budget), leads to the right quality;
— to work in team;
— to effectively communicate with the different stakeholders involved in a project;
— to understand the business needs and goals underlying a project and to evaluate the trade-off between different solutions;
— to effectively present the proposed solutions providing rationale behind them.

Several issues are involved in those learning objectives and simply lecturing and discussing software development phases, methodologies, tools, and techniques is not sufficient to provide the suitable preparation for a professional career. Real-life experiences are needed to give students the opportunity to be exposed to and understand the many organizational, human, and methodological issues involved in an IT solution development, to recognize the importance of soft skills and discover how work on themselves to acquire them. Without doubts team projects should be a must of any Computing curriculum, since real-life projects are never the endeavour of a single individual but require the effort of several people. But how those experiences should be organized to be effective?

In our opinion, it is important that students work on a real-life project coming from a real need of a company or an organization to have a better understanding of real-world problems. This helps them understand the context in which IT solution operates and experience the importance of good communication skills needed to identify the scope and elicit the requirements that best meet organization business needs and satisfy constraints. IT solutions pervasiveness require that IT professionals are quite flexible and able to quickly understand and acquire the language characterizing each application domain. Moreover, it is important that they are exposed to broader systems issues, since computer systems are often part of a larger system, including mechanical and/or electrical systems, each one characterized by specific features and constraints. Thus, it is important to be able to effectively communicate and collaborate within the multidisciplinary team in charge to design and develop those complex systems, making trade-offs between different solutions. Those teams are made of engineers from different disciplines (mechanical engineering, electrical engineering, management engineering, software engineering...) each one employing a different technical language. The communication issues can be further stressed by the fact that those teams can also be international and intercultural, thus motivating the need for an international dimension of Computing curricula. It is widely recognized that any science and technology subject is inherently international since they involve international collaborations between groups of scientists/practitioners located around the world. This is especially true for Computing due to some specific features of the field, such as global employability, IT systems' global market, the increased globalization of software development, the widespread use of offshoring, etc. These features and appealing opportunities pose new challenges and issues to IT practitioners also because Computing (differently from other sciences) involves a human and social dimension. In particular, to operate in this global environment we need to prepare students, so that they are able to:

- understand that IT systems need to be tailored to the local end users characterized by specific cultures;
- understand the cultural frame of the organization where they are employed;
- work in multicultural teams;
- effectively communicate also in foreign languages with internal and external stakeholders of an IT project [Fuller 2005].

Living an international mobility experience is widely recognized as an opportunity for personal and professional growth, with special regard to the acquisition of a better awareness and understanding of cultural aspects and of communication skills in foreign languages.

The need to add an international dimension to Computing curricula is driven not only by some specific features of the field but also by institutional, economic, and political agendas.

### 1.2 Needs for internationalization

Due to the external as well as internal demands Higher Education Institutions are motivated to increase and promote educational internationalization activities. Indeed, internationalization is widely recognized as one of the key indicators for Institutions quality assessment [Altbach and Knight, 2007] [de Wit, 2010] and in some cases it is also used for determining the Government funds (see e.g. [Italian Law L 43/2005]). So, internationalization is the mainstream of institutional strategic development and part of the public statement of the mission and profile of many Higher Education Institutions. Notable in this sense is the statement of University of South Australia where not only graduate international perspectives (as a professional and as a citizen) is reported as one of the seven qualities the institution intends to pursue but also a set of useful indicators are identified which serve as a guide to its development [University of South Australia].

Internationalization is one of the key elements of the European Higher education reform playing a crucial role for the development of European Higher Education Area. So it has a high priority on the agendas of European Commission and national governments.

In the last decades, internationalization is evolved and enlarged and now the term encompasses various forms of international cooperation among institutions, student and staff mobility across national borders, and cross-border transnational delivery of programmes. As a matter of fact, several programmes have been developed to meet the strategic European objectives. These range from the need to harmonize higher education systems to improve quality; to meet the needs of the social, economic, and technological development of the knowledge society; to increase the global competitiveness of European higher education. The funded programmes range from the well-known and widely adopted ones (e.g. students and staff mobility), to more emerging ones (such as mobility for Placement and Intensive Programme) to niche programmes (such as Erasmus Mundus). All the above programmes and many others have been encompassed in the new integrated programme Erasmus+ that aims to boost skills and employability, as well as modernizing Education. The EU's commitment to investing in these areas is reflected by the increasing budget assigned for the period 2014-2020 (about a 40% increase compared to current spending levels).

Mobility is the crucial feature underlying the internationalization vision, it "is considered to be the most important reason for making internationalization a priority and is identified as the fastest growing aspect of internationalization" [Knight 2003]. Indeed, it not only allows students to take advantage of different higher education systems, but also "offers students the opportunity to learn the skills needed to escape their wells and learn to swim in the ocean beyond" [Leask 2007] by allowing for intercultural dialogue, exchange and knowledge. Indeed, it is an important experience to qualify students as open-minded future professionals able to consciously and responsibly live in a cohesive society.

### 1.3 Content and organization

In this paper we present the design, objectives, experiences, and results of two developed editions of an internationalization project funded by the European Commission in the context of the Erasmus Intensive Programme (IP, for short) highlighting how the project contributes to reach the key learning outcomes mentioned in the previous sections and to meet the goals of the above internationalization vision. The Erasmus IP was part of the Life Long Learning programme the European Union programme for education and training for the period 2007-2013.

An IP is a short programme of study that brings together university students and staff from no less of three countries in order to encourage efficient and multinational teaching of specialist topics, which might otherwise not be taught at all, or only in a very restricted number of universities. An IP requires transnational academic co-ordination in order to organize a course during at least 10 continuous working days. Funding can cover organizational costs, transnational travel and subsistence costs of teaching staff and students directly related to the participation of the IP.

In this context, the present IP was organized in order to provide some Computing skills and additional learning outcomes in the context of IT applied to industrial processes. The IP put together students and lectures with different scientific and cultural knowledge, which participate in a short (two weeks) but intensive project that suitably integrates an academic and social programme. A first edition, named Simulation and Virtual Reality in Robotics for Industrial Assembly Processes Intensive Programme (SAVRO) ran in the period 2008-2010 involving three Institutions from three European countries. The second edition, Human-Machine Interaction (HUMAIN) ran from 2011 until 2013 with an additional partner.

The aim of the paper is to illustrate some best practices that characterize the organization of the IPs so that other Institutions can take advantage of them for the development of other similar international experiences. Moreover, we will present some lessons learned from our experience that allow us to make some general recommendations and suggestions on how to devise Computing academic curricula.

The paper is organized as follows. In next section the antecedences, objectives and development of both editions of the IP are introduced. Their organization is explained in section 3, including implementation problems. The fourth section briefly shows the projects carried out by the students and their results. The evaluation of the IPs carried out by means of surveys to the participants is presented in section 5. The paper ends with some concluding remarks and suggestions.

## 2. THE SAVRO AND HUMAIN INTENSIVE PROGRAMMES

### 2.1 Antecedences

The origin of the Intensive Programmes comes from the long collaboration between some of the partners in mobility programmes. For example, since 1997 there has been a Socrates-Erasmus agreement for the exchange of teachers and students between the *Facultad de Informática*, from the *Universitat Politècnica de València* (UPV) in Spain, and the *Fachbereich Maschinenbau und Verfahrenstechnik* of the *Technische Universität Kaiserslautern*, in Germany. In 2003, a joint seminar was organized in Valencia, sharing academic activities mainly in the domain of robot programming and simulation.

During the 2004-05, 2005-06 and 2006-07 courses, the cooperation was increased with a new academic project, called European Team Seminar on Robotics, funded by the *Verband der Pfälzischen Metall- und Elektroindustrie* (*Pfalzmetall*), a German agency that promotes the mechanical and electrical industry in the Pfalz region. This innovative project consisted of the realization of an industrial project for a German company located in the surroundings of Kaiserslautern. This institution promoted the project to improve the multinational and multidisciplinary education of the engineers of the region.

The students had to organize themselves and distribute the work between the different participating groups. Initial assessment, subdivision into sub-projects, scheduling and communication (via-Internet, using English as the official language) between both groups were different duties that students had to undertake during the project. A total of 34 German students from the courses of mechanical engineering and electrical engineering and 15 Spanish students from the Robotics and CAD courses of the Informatics Degree participated [Mellado et al. 2005], [Mellado et al. 2006], [Mellado et al. 2007]. Both teams (with skills that complement each other) were working on a real industrial project given by a company during a whole term (4 months). Finally, both groups presented the work to the company in a joint presentation. Therefore, the cooperation and contact between the different teams were a key factor to the successful completion of the task. During these 3 years the following companies have participated in the project: *Sensus Metering Systems GMBH* (www.sensus.com), *Keiper* (www.keiper.de) and *Terex-Demag* (www.terex-demag.de).

The Intensive Programme that will be presented in next sections started from the experiences gained in these projects and the extension with new areas of work and new participants. For the first edition of the IP, called SAVRO [Vendrell et al. 2009], a third participant was joined to the consortium, the *Facoltà di Scienze Matematiche, Fisiche e Naturali* of the *Università degli Studi di Salerno* (Italy), which contributed with their knowledge of Virtual Reality and their use in assembly processes. Finally, for the second edition called HUMAIN, the *UFR Sciences Exactes et Naturelles* and *the IUT Reims-Châlons-Charleville* from the *Université de Reims Champagne-Ardenne* (France) became the fourth member of the consortium.

### 2.2 Objectives

The described IP is an educational project whose main task is the training of the participant students on the necessary skills for the development of a solution for a practical case applied to a real industrial environment. It also involves international and interdisciplinary relations, offering the students taking part in the program a valuable and distinctive feature to add to their curriculum vitae. With the aim of exposing students to broader systems issues for helping them understand the context in which software operates, often as a part of more complex systems including mechanical and electrical parts, we focused on standard robotics and industrial automation problems. In particular, the aim of the IP was to teach the students a methodology to address problems in those domains by the usage of simulation and virtual reality software in real processes to define graphical operation interface.

The project aimed to create a team, made up of groups of students from Computing, Electrical and Mechanical Engineering and Automation Engineering degrees, so that combining the knowledge of the specific groups, results in a global team that has the required skills to solve the task of analysing and optimizing industrial assembly processes and designing their human interfaces.

The activities in this IP included the analysis of a real industry case, the learning to use software tools useful for automation, the learning about assembly processes, the design of graphical user interfaces and the development of an industrial oriented team project.

The IP operative objectives were to improve the quality and increase the amount of cooperation between academic institutions and companies as well as to promote the development of innovative educational techniques for graduate level courses and export them from one participating country to others.

The IP intended to achieve the following goals:

— Create an academic project that includes an analysis and research task on a representative case of an industrial application, emphasizing the use of specific software tools that are helpful in that task.
— Enhancing the student's analysis, design and programming skills of industrial applications with the use of specific software tools for simulation and design of the interaction of the operator with the system.
— Teaching the students to work in multinational and multidisciplinary groups.
— Enhancing the student's knowledge of other disciplines that share the same academic objective.
— Encourage group learning in an international multilingual environment.

— Promote soft skills to the students, gaining experiences in cultural, communication and language integration,
— Establishing a stable international network with regards to the teaching of robot automation applied to industrial processes.

The final objective of this IP was to create a context that integrates academically and culturally students and professors of different EU countries within a common area of teaching and research. Also, participating professors had the opportunity to share experiences and find contacts that give them a new perspective in the areas of teaching and research.

In agreement with the objectives specified in the Erasmus Charter of each participant institution, the IP established as priorities the integration of participant students in an international environment, increasing the mobility and use of foreign languages, in this case English, non-official language in all participant countries. Furthermore, one of the objectives of the IP is to assess the students' workload while working on an industrial application practical case, as a method to evaluate the concepts learnt in the courses related to the project.

### 2.3 Development

The course was organized with a similar structure for all editions but considering a different industrial project for each one. Each year course was focused on the development of a different team project, in the area of industrial process automation, with real requirements formulated by an industrial partner.

The IP included two weeks of international mobility at a partner institution, with lectures in English and the development of a team project, but also pushing the international integration with social and cultural activities.

The project aimed to provide the necessary bridge between academia and the professional environment, with the active participation of industrial partners that suggested actual problems and supported in the assessment of the proposed solutions. Moreover, the project reflected a usual situation in the labour market where complex systems need the contribution of different disciplines and areas of expertise to obtain profitable solutions and to adapt to international requirements. The rule of the industrial partner was essential to success in this goal. Six different companies participated in this experience, coming mainly from the automotive sector (i.e. Fiat) and from the food and beverage sector (as the industrial bakery *Dulcesol* or the champagne producer *Nicolas Feuillatte*), selected by the local university partner.

Activities in the project included the analysis of a real industry case, the learning of automation software tools, the design of graphical user interfaces, the development of an industrial oriented project, including its programming, and the presentation of the proposed solution by illustrating both, the expected throughput of the solution and the impact in terms of efficiency and costs for the company.

The teaching project covered different general and specific competences and learning outcomes, giving the students additional skills that complement their basic knowledge. These skills are essential requirements in the dynamic professional environment that the students have to face when they get the degree. In particular, the general transversal soft skills covered by the IP courses were:

— Teamwork, leadership, communications, collaboration and organizational skills.
— Use of a foreign language (English).
— Integrating and working in multidisciplinary teams with different background.

— Ability of understanding the organizational processes for which the software is developed and the business goals of the organization where they work to better identify the right problem solutions.
— Ability to work in a multicultural context as required in a global distributed software development project and for global employability.

## 3. THE ORGANIZATION OF THE INTENSIVE PROGRAMME

### 3.1 Management of the courses

Management of the IP was intended as a joint venture, supervised by the coordinating entity. In order to administer the contribution of all participating institutions, we used a CMS Webct tool based on *Sakai*, an open source learning management system (www.sakaiproject.org), and expressly designed web sites for the IP, with restricted access to share information between the different members of the consortium. Social applications, as Facebook, were also used to facilitate the communication between participants.

The use of the learning platform allowed all participants in the course, lecturers and students, to interact, in a first stage, during the period previous to the course and preparing documentation, sharing teaching materials. Then, during the course itself, the platform was used for learning purposes. Hence, tasks and evaluation activities were performed by means of this resource, making the teaching-learning process easy, and giving the students a dynamic perspective of this intensive course.

The development of the IP was integrated as a novel teaching experience, with the active participation of the students, according to innovative education programmes that are being done in the different institutions, as stated in the Bologna Process [Heitmann, 2006].

There were three clearly separated phases in the IP:

(1) The first phase was centred on the preparation and organization of the intensive course. It consists of two main points. First, the academic preparation of the course, including the design and the contents of the lectures, the definition of the teaching methods, the editing and production of the material, its diffusion, etc. Also, during this phase the IP was publicized through a corporative image (web site, logo, documentation and presentations, flyers, posters, etc.), in order to select the students. All partners of the IP, guided by the coordinating University, who has a major part in the design of the publicity material, were participating in the development of this task. Furthermore, as a second point to be covered in this first phase, the search for participating industries in the IP was done. Subsequently, some contacts were established with companies interested in collaborating in the IP, so a practical problem about optimization can be found and designed. The local partner was in charge to do this task previously to the start of the IP.

(2) The second phase was basically the implementation of the course. Lectures and additional cultural and sportive activities were organized, as explained in next section. The project development, visit to the company and the presentation of the project results were also included in this phase.

(3) The third phase, after the course had taken place, consists of an assessment of the course and the publication of results. Also, in this phase, administrative works, such as budget justification and renewal application were done.

Figure 1 shows a diagram summarizing the organization of the IP course.

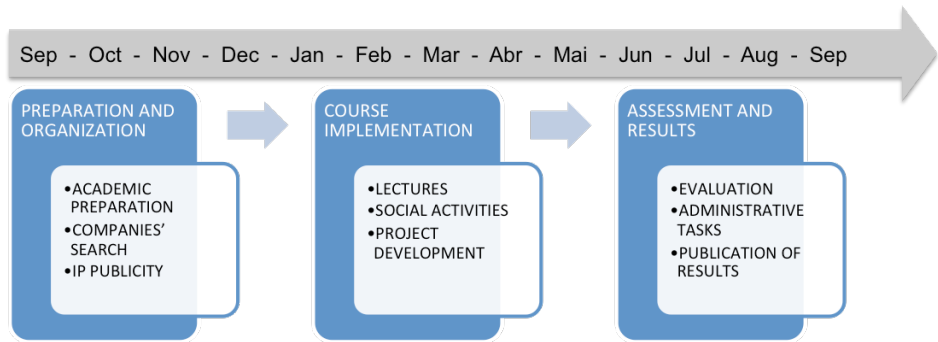

Figure 1. Organization of the IP course.

The supervision of the IP was done by the coordinating University, managed by its International Relation Office. This supervision consisted in guaranteeing the adequate publicity of the program among the students of each institution, management of the web site associated to the programme, and control of the different actions included in the programme. Also, the coordinating partner was in charge of giving attending certificates to the students and professors participating in each edition of the IP. Another responsibility of the coordinating entity was the organization of an annual meeting with the rest of the institutions to establish the schedule, agenda and contents of each edition of the programme, the request for renewal with new improvements, even considering new possible institution members.

The organization of the IP was done in a structured manner, having a person in charge for each participant institution. Every participant professor in the IP is a teacher in his/her university, in different subjects related to the IP main topic. This aspect guaranteed the quality of the contents to be taught in the project, as well as the perspective of having stable teaching resources in the IP.

It is also important to mention all the dissemination activities involved in the IP. These activities are part of the organization of the course and can be summarized in the next list:

— A corporate image has been designed, including a logo to be used to identify all teaching materials in the IP.
— Some diptychs and posters have been designed in order to publicize the IP.
— An official web site for the IP (as mentioned above), as the main site for all participants in the course. This site includes relevant information, news, a download section, links to the course's site in the learning platform and link to social platform for the group (the first two years a blog was used, while the rest years, the consortium created a Facebook site). Both social platforms were intended to be used mainly by the students as a meeting point for non-academic activities.
— A short paper describing the IP course was included in some editions of a national magazine, edited by the OAPEE, the Spanish institution in charge of organizing all activities involved in the Long-Life Programme of the European Commission [OAPEE 2008].
— A video report, including some interviews to students and teachers, was prepared for each edition of the course, in the way of a making-of video. The video was edited for broadcast on the UPV local television.

The courses were hosted each year in a different country. In the total of the two editions, the project involved almost 150 students (77 in SAVRO and 72 in HUMAIN) from 10 European countries.

### 3.2 Lecturing and other activities

The IP course was scheduled in 10 workdays, during 2 consecutive weeks (Monday to Friday). During the course, the students combine lectures and lab seminars with non-assisted work. This autonomous work was performed in labs and is part of the teaching schedule of the course in accordance with the European Credit Transfer System requirements.

Therefore, the teaching method combined theoretical lectures and practical seminars with teamwork for project development, according to the following structure:

a) During the first week, the students attended to lectures on topics required for project development and lab seminars for training with practical software tools.

   According to the project for a year, different subjects were taught to the students during the course. Some of the topics that have been covered along the six years are the following:

   — Robotics

   — 3D graphics and virtual reality

   — User interfaces design

   — Human-machine interaction – general aspects

   — Graphical user interfaces – practical aspects

   — Augmented reality & haptic devices

   — Industrial interfaces – applications with SCADA systems

   — Design of assembly processes and lines

   In addition, lab seminars were organized in practical sessions for teaching about the software tools required for project development. For example, seminars have covered:

   — Modelling and simulation in robotics

   — Programming a robot simulator

   — Modelling with the SW interface tool

   — Programming the SW interface tool

   — Design of industrial interfaces

   The students had free access to a computer laboratory to develop some exercise to practice with the software.

b) The students, organized in work teams, had to develop a project along the second week and make a final presentation of their results. The project was defined during a visit to a company and discussed with the company engineers. Typical agenda of the visit included:

   i. Company welcome with a presentation of the company history, evolution, current figures, and products

   ii. Visit to production lines

   iii. Explanation of the problem to be covered in the project

   iv. Data collection by students

v. Discussion with production engineers of the company
   vi. Acknowledgment to the company and departure

In this phase, the students assume a learning method of active and coordinated learning.

c) In parallel to all these lecturing activities, additional actions were arranged in a social programme, including sport competitions, visit to main local sightsee places and cultural heritage locations.

**3.3 Skills and learning outcomes**

One of the main purposes of the course was to set up the basis of a multidisciplinary topic in the area of Computing and Industrial Automation. This topic, which can be intended as a specialization, covers different general and specific competences and learning outcomes, giving the students additional skills that complement basic ones previously acquired in their degrees. This specialized teaching objective is a crucial aspect in the dynamic professional environment that the students have to face when they get a job [Kandra et al., 2011].

These are the general transversal competences covered by the IP:

— Teamwork.
— Use of a foreign language.
— Integrating and working in multidisciplinary teams.
— Working and leading projects in the Computer Science context.
— Being pro-active and self-organized when facing real problems.
— Communication in different ways (oral and written) the results of the proposed solution.

Related to specific skills, next is a list of learning outcomes in different areas, covered by the IP. These learning outcomes have been considered according to the ACM-AIS-IEEE Computing Curricula recommendations [ACM-AIS-IEEE 2005].

— Develop solutions to programming problems.
— Do systems programming.
— Create a software user interface.
— Define information system requirements.
— Implement intelligent systems.

As real industrial problems have to be faced, comprehension of a system was considered. In this sense, skills and competences from the Systems Engineering field have been also considered in the IP [SEBoK v. 1.3 2014]. Next, some basic learning outcomes related to this approach are listed:

— Analyse a system, describing its basic behaviour, properties and functions.
— Describe the structure of a system.
— Identify the elements of a system.

In addition to these general and specific skills, and as a natural extension from the Systems Engineering competences, the course covers new specific learning outcomes that must be considered in a context where Computing has to be combined with other main topics, concretely Industrial Processes. In this sense, this international Erasmus IP tries to cover these specific learning outcomes:

— Analysis, modelling and simulation of industrial processes.

—Define requirements and input/output information for industrial processes.

—Develop solutions to industrial problems using computer systems.

—Design and program industrial processes considering resources and layouts.

In summary, this IP combines general and specific learning outcomes with the final purpose to give specific skills to students coming from different degrees in order they can face current challenges and issues of a dynamic professional environment.

**3.4 Implementation issues**

Organizing an international course with the participation of lecturers and students from four different countries was not an easy task. Many efforts in coordination had to be done in order to success in its implementation. For example, an IP required moving all the students to one country during at least two weeks. Due to the academic calendar of each country involved in our case (France, Germany, Italy and Spain) it was difficult to find two weeks that suited well to everybody. It was not only a problem of National and local holydays. The main problem was related to exam periods. As each university partner had its own teaching and examination schedule, course dates had been always the first problem to solve. A long meeting and some virtual meetings were usually required to get a common solution.

Next problem to solve was student selection. Nowadays, students have so many alternative options for mobility and internships that high motivation was required to attract some to participate in an IP. To take advance to the mobility competitions, talks about the IP opportunities and advantages were organized in each university to motivate student participation. It was a hard task, but all the vacancies were filled every year. This was a success mainly for the commitment of all involved partners.

One of the main problems, even for student participation, was that Erasmus IP funding did not cover all the expenses and only 80% of the budget is paid in advance. From the final amount, a 75% of actual travel cost was covered and the budget for accommodation and daily expenses was really very low. For example, each student had a total of 376.32€ for accommodation and stay during a week in Germany, that is, 53.76€ per day (insurance and local transport included). It was very difficult to find accommodation with so low rates. In this context, the partners had to find extra funding from each university or grants from companies to increase the budget for the students to avoid withdraw.

Local organization was always complicated. For lessons and lab seminars, classrooms and laboratories had to be reserved and sometimes were not so easy to get them for full-time for two weeks. Other university facilities were also required, as some space in university restaurants or special sport facilities. Laboratories had usually the operating system and all software of the computers in local language (French, German, Italian or Spanish) while the official language of the course is English. So, when possible, computers had to be set up for the course. As an anecdote, computer keyboards were different according to the host country.

One of the most important aspects or our intensive programmes was the participation of a company to propose an industrial project. It had been very difficult to convince companies to participate, as they did not get any advantage and even some considered a loss of time. Therefore, it was another task for the lecturers who took advantage of their personal and professional contacts to persuade them for their participation. In addition, the industrial project had to be easy enough to be analysed for one week but, at the same time, hard enough to be a motivating challenge for the students. The quality of the partners and the voluntary effort that their staff dedicated to contribute to the success of the IP every year deserved a great acknowledgment.

As punctual complications that were to be faced along the first years of the IP, there were a couple of student withdraws few weeks before departing. They were caused in one case by a job opportunity and in the other one by a personal situation. Since these two cases, students had to sign a commitment letter to be accepted in the course. Finally, there was a student that had a health problem during the IP, but the insurance contracted for all the participants each year take care of the entire situation, including the repatriation to home.

## 4. RESULTS: PROJECTS DEVELOPED BY THE STUDENTS

Prior to the course, the local coordinator searches for a company partner and agrees an industrial project with the company production engineers. These engineers participate in the IP course hosting a visit to the company. During the visit, the students can learn about the production process of the company. The visit ends with a discussion about the project requirements with the participation of the students with the company engineers and the lecturers.

During the second week of the course, work teams are established with the aim of doing the project that must be presented the last day of the course to the lecturers and the company engineers with the solution proposal developed by the students.

The teams are organized with students from different nationalities to strengthen cultural interchange and reinforce the internationalization aspects of the IP. It is very profitable for any student to learn how a problem is dealt with from different idiosyncrasies.

At the same time, each student in a team has a different background, as there are students from computing degrees, computer science degrees, mechanical/electrical engineering degrees and automation engineering degrees. The conjunction of backgrounds facilitates the correct development of the project, as the proposed projects cover aspects that hardly can be solved from a unique profile of education. This combination of backgrounds permits to attack the project from different points of view and the learning process of the students is enriched at the same time, as they are working in the same way as in many jobs of their professional future.

The last day of the course every team must present its results as a solution proposal to a board of examiners with lecturers and company engineers as members. Usually, the students have to offer a detailed technical solution with the simulation of the proposal and a draft design of the system interface for the operator.

In order to motivate the students, a best project competition is organized in such a way that the board of examiners selects the team with the best solution considering the potential use of the solution in the company, its originality and the quality of the presentation. This team receives a special certificate for this achievement.

The following two sections include a brief description of the three projects carried out in each of the IP editions, SAVRO and HUMAIN.

### 4.1 SAVRO projects

The projects covered in SAVRO edition were all related to automotive industry, with the participation of a scrap company, a research centre of *FIAT Group* and a manufacturer of auxiliary products as car element provider.

The first year of SAVRO was hosted in Valencia in April 2008. That year, a research project carried out at the Instituto ai2 at UPV was focused on the disassembly process of cars in the post usage life phase. Demands of society (environmental concerns, climate change, etc.) and the legal frame (European Directive 2000/53/CE, national and local laws, etc.) are forcing manufacturing industry to reuse and recycle products at the end of their life for ecologic and economic reasons. Therefore, in order to fulfil these demands, the design of automatic disassembly cells for end-of-life products is an imperative issue, as this process is heavily hand-based. With the help of the SME company *Desguace Malvarrosa* (www.desguacemalvarrosa.es), a study of this project was explained to the students and they have to work in teams on different aspects that they had to select, such as:

— Develop disassembly workflow and layout

— Proposal of a deliver system of the car

— Design and operation of the proper tools

— Selection and information process of the sensors

— Extraction of car battery

— Extraction of residual fuel

— Simulation of processes involved

In Salerno, in March 2009, the second year of SAVRO was held. The industrial project was proposed from *Elasis*, one of the principal private sector research centres on automotive industry in Europe and member of *FIAT Group*. The project was about the automation of an operation in the production process of a car which had to be done manually by that time. More precisely, two workmen had to carry a doorframe from a framework and position it in a final framework according to its shape, changing the orientation during the task. To automate this process and make it more efficient, the students had to propose an automation process based on robots to accomplish and repeat the process with some requirements as cycle time or cell size and with an adequate precision in each repetition. The new process must be graphically simulated to show the possibilities of implantation in a real production line. After one week of project development, the students presented their proposals.

The last year of SAVRO, the IP was held in Kaiserslautern during two weeks of February and March 2010. The company selected was *Mann+Hummel GmbH* (www.mann-hummel.com), a manufacturer of industrial filters. The project was about the automation of the production process of a filter for automotive industry without plastic core which had parts done manually. More precisely, a workman had to glue the filter material in a coiling machine and carry it to a drying machine. Because of the features and situation of the wrapper and the drying table, the process lasts 2 or 3 additional minutes for sealing the side of the filter. The implantation of a robot to substitute the operator was considered by all the projects except one, which proposes a multi-track transportation system as the cheapest solution. The team won the best project recognition, and the company considered its proposal for real implementation.

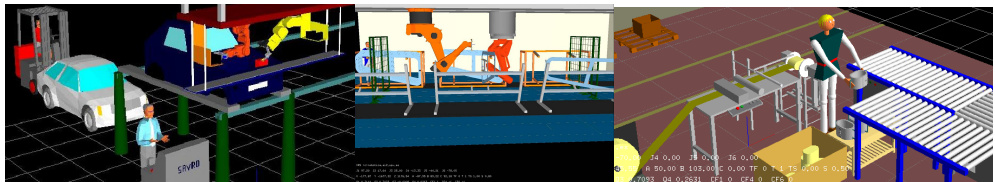
Figure 2. Screenshots of the simulation in the three SAVRO projects

### 4.2 HUMAIN projects

In the second IP edition, HUMAIN, all the projects were related to food industry, with the participation of three producer companies: a bakery and pastry maker, a public company for milk bottling and distribution, and finally a champagne producer company.

The first year of HUMAIN was hosted in Valencia in April 2011. The company selected for collaborating in the IP was *Dulcesol* (www.dulcesol.com), the Spanish bakery leader settled in Gandía, 60 km away from Valencia. Its most important activity is the production and marketing of pastry, cake and bread products. The project was to give ideas for the automation of the primary packaging of the product "*Tarta de anís*", a round flat cake which is packed in stacks with several packages' formats. The student project was a real challenge, as there were requirements on layout, product fragility, and high demands on reduction of cycle time. Some of the solutions were highly valued by the Production Director in *Dulcesol*.

The following year, *Centrale del Latte di Salerno* (www.centralelatte.sa.it) a company responsible for the production and sale of fresh milk and dairy products, was the company partner in Salerno, Italy for the IP during two weeks in April 2012. The production engineer showed a project in its initial state as they were waiting for funding. The project consisted on the preparation of boxes coming from tracks on stacks of six boxes, for automatic cleaning and storing process. There were a high variety of solutions proposed by the teams, from application of industrial robot-arms to avoiding them for the cost, including the use of a couple of mobile robots.

The last year of HUMAIN, the IP was hosted in Reims in March 2013. The company selected was *Nicolas Feuillatte* (www.nicolas-feuillatte.com), a well-known centre for sparkling Champagne wine production and the largest cooperative union in Champagne region. The project was about the automatic storage of thousands of bottles of champagne in cellars. The students have to find solutions which must be efficient, economic and ergonomic.

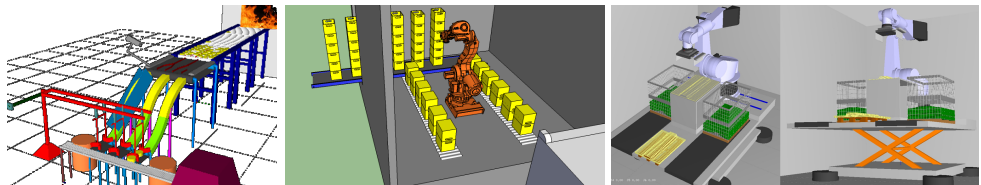
Figure 3. Screenshots of the simulation in the three HUMAIN projects

### 5. COURSE EVALUATION

The assessment of the IP was done with an anonymous survey that all the participants must complete at the end of every course. This survey, included as appendix in this paper, asked the participants about the different features of that IP, the contents and the organization. The results are taken into account when planning the subsequent IP courses.

Next, some results drawn from the surveys are listed:

— The contents of the course satisfied the students' expectation, in the sense that they experienced a multidisciplinary approach to real industrial problems. Between 90% and 95% of students (depending on the IP course) consider satisfactory or very satisfactory the experience.

— The academic organization of the course was well considered (see the charts below). Students thought that the distribution of theoretical sessions during the first week, and teamwork sessions during the second week was appropriate to the final purpose of the course.
— The students (around 90%) considered that working in a real project proposal, derived from a visit to a company partner of the consortium, was an added value to their academic curricula.
— Social and cultural activities organized in parallel to the course were very highly regarded, so the students thought that these actions advocate real integration among them.

Next, some charts about the survey results are included from some years of HUMAIN edition of the IP, showing the high acceptance of the IP course among the students.

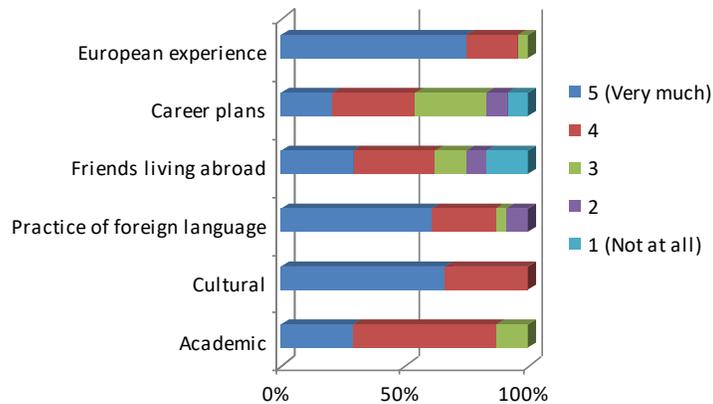

Chart 1. Factors that motivated to participate in the course (HUMAIN IP 1st year).

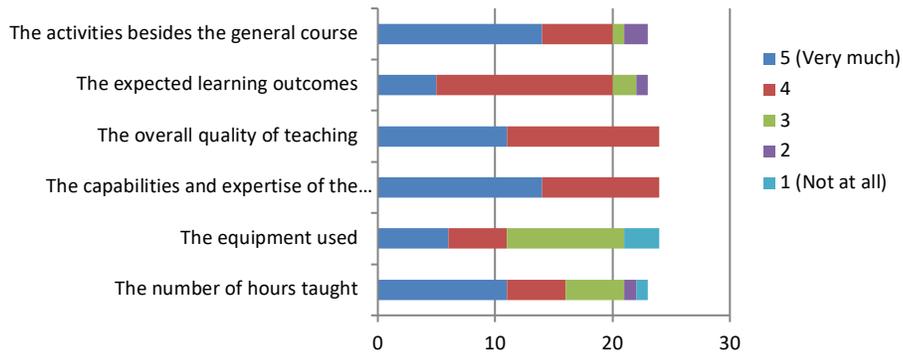

Chart 2. Satisfaction with academic activities and pedagogical aspects (HUMAIN IP 3d year).

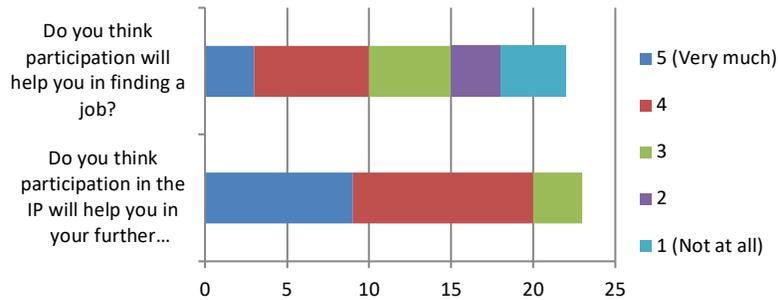

Chart 3. Opinion about the impact on academic and professional issues (HUMAIN IP 3d year).

At the end of the survey, students were allowed to provide feedback in terms of their thoughts regarding the IP. The following is a sample of written feedback that were received and that represent a great synthesis of our work and students experience.

— "Learning is experience. Everything else is just information" (Albert Einstein). Pleasurable experiences not just information is exactly what we bear in our hearts, study experiences and life experiences. What do they have in common? A lot, considering they both entail a continuous learning to new achievements and allow us to ever bring ourselves into play, for better or worse, and to know ourselves and others.
— I consider our experience as an opportunity, a rendezvous for cultural socialization that should be part of all formative pathways.
— We lived so close to each other for two memorable weeks, and we had to tackle various people from different countries, involving us personally and professionally.
— We have made new friends and got to know old ones better. I would love to keep in touch with you all.
— Life is so unpredictable; everything can change so quickly that we could be even unaware of what we are losing, the sense of emotions and experiences.
— Who knows, maybe a day we'll work together.
— Let us live and learn, let us know the world, that is not only what we have inside, but all is around us and we often judge as just information!
— I wish to express my gratitude and to thank you all and the project coordinators for giving us the chance to share such a gratifying experience.

## 6. CONCLUSIONS

The European Commission has offered several funding programmes for internationalization in its Life Long Learning programme. They differed with respect to the pursued objectives and to the resources and efforts that Institutions had to put in action to devise and manage them (e.g. extraordinary organization efforts were needed for an Erasmus Mundus Master). The Erasmus Intensive Programme (IP) has been a very profitable programme that was definitively cost-effective in terms of the balance between efforts necessary and results that can be obtained. With the launch of the new Erasmus+ programme, Intensive Study Programmes are part of Higher Education Strategic Partnerships whose aim is to support the development, transfer and/or implementation of innovative practices making use of IT and Open

Educational Resources as well as the implementation of joint initiatives reinforcing cross-sectorial cooperation, and exchanges of experience at European level.

In this paper we have described the main features and results of an Erasmus Intensive Programme that ran along six years between four European universities. Its objectives, development, management and results have been explained, as well as some implementation problems and how we solved them.

Along the paper, we have highlighted some best practices that can be exploited by others organizing similar active teaching activities. One of the main challenges in a dynamic professional environment is about giving solutions to multidisciplinary industrial problems. Computing skills have to be combined with other abilities in order to get new learning outcomes and competences to allow new professionals facing difficult tasks under new approaches. The IP were focussed on teaching the students this new approach and how to affront "real-life" experience in the form of industrial projects given by actual companies. In this sense, a specific organization of the course was designed. So, combining theoretical lectures with practical sessions and autonomous teamwork, including roll assignment to students, and managing different mixed teams in an international environment moved to get the objective of the course in the students.

The success of the IP was reflected in the results of the surveys used for assessment. In this context, the students considered that the model established for this IP is perfectly tailored to the pursued objectives. The IP synergistically integrated in the same programme a social, cultural and didactic oriented project. The experience carried out allowed creating a kind of "IP Community" as a premise for future working synergies by the members.

The IP experience allows us to make also some suggestions and recommendations for devising Computing curricula. Indeed, students have appreciated the active learning approach characterizing the IP and mainly the fact that the project proposal came from a real need and requirement of the industrial world. In our opinion, this approach could be usefully extended to devise Computing curricula to increase students' motivation and make them more convinced on the value of the chosen academic curriculum and on the competences and skills it will provide them to cope with the labour market. This is also coherent with the proposals funded by the National Science Foundation to Revitalize Undergraduate Computing Education (CPATH). The selected curricula are usually centred on the development of solutions for specific real problems (coming from the industrial world).

Multidisciplinary work is another feature characterizing the IP and much appreciated by the students. It reflects a usual situation in the labour market. Indeed, often software and computing systems are used to solve a problem in a specific context where they are only a part of a more complex system. So, we should get our Computing students more used to work in a multidisciplinary context. It is not necessary to create an international project to have a multidisciplinary experience, since synergies could be realized with other components and areas of the University. In Italy, the national law requires several ECTS associated to courses in interdisciplinary areas. However, this requirement is usually covered by simply offering non-Computing courses. In our opinion, to better meet the goal, Computing curricula should include the realization of a project in a multidisciplinary team (including students from different areas) as a complement to these courses or a way to cover the required ECTS.

**APPENDIX**

Next pages show a sample of a survey to be filled by the students enrolled in the IP courses.

| 1. Student details |   |
|---|---|
| Name (not compulsory): | |
| Home higher education institution: | |
| Erasmus ID code of the home institution: | |
| Subject area of your degree/major (using the nomenclature ISCED 97): | |

| 2. Identification of IP and motivation | |
|---|---|
| Title of the IP (to be prefilled by the organiser): | |
| Location of the IP (to be prefilled by the organiser): | |
| Host institution (to be prefilled by the organiser): | |
| Erasmus ID code of the host institution: | |
| Dates of the IP (to be prefilled by the organiser): | |
| How satisfied were you with the duration of the IP: | |
|    Scale 1-5: 1=not at all, 5=very much | 1 – 2 – 3 – 4 – 5 |
| How satisfied were you with the dates of the IP? | |
|    Scale 1-5: 1=not at all, 5=very much | 1 – 2 – 3 – 4 – 5 |
| Which were the factors which motivated you to participate? | |
|    Scale 1-5: 1=not at all, 5=very much | |
| Academic | 1 – 2 – 3 – 4 – 5 |
| Cultural | 1 – 2 – 3 – 4 – 5 |
| Practice of foreign language | 1 – 2 – 3 – 4 – 5 |
| Friends living abroad | 1 – 2 – 3 – 4 – 5 |
| Career plans | 1 – 2 – 3 – 4 – 5 |
| European experience | 1 – 2 – 3 – 4 – 5 |
| Other (please specify): …………………. | |

| 3. Information and support | |
|---|---|
| Where did you hear about the IP? | |
| Home institution | Yes / No |
| Host institution | Yes / No |
| Other students | Yes / No |
| Former participants | Yes / No |
| Internet | Yes / Other |
| | (specify) : ……………………….. |
| Did you receive adequate support from your home institution and from the host institution before and during the IP? | |
|    Scale 1-5: 1=poor/negative, 5=excellent | |
| Home institution | 1 – 2 – 3 – 4 – 5 |
| Host institution | 1 – 2 – 3 – 4 – 5 |

| 4. Accommodation and infrastructure | |
|---|---|
| Type of your accommodation in the host country: | |
| University accommodation | Yes / No |
| Apartment or house together with other students | Yes / No |
| Private housing | Yes / Other |
| | (specify): ……………………….. |
| Were you satisfied with your accommodation? | |
|    Scale 1-5: 1=not at all, 5=very much | 1 – 2 – 3 – 4 – 5 |

| 5. Recognition |
|---|

Will you gain recognition for your IP by your home institution?
                 Yes / No / I don't know yet
If yes, how will it be recognised?
                     ECTS
                 Diploma supplement
                 Other (please specify):
                 ……………………….
Did you encounter any problems concerning recognition of your IP?
 Scale 1-5: 1=not at all, 5=very much        1 – 2 – 3 – 4 – 5
                    Please specify:
                  ……………………….

| 6. Costs |
|---|

| Total approximate personal contribution to the costs of the IP (EUR): | |
|---|---|
| What kind of costs did you need to contribute to? | |
| Travel to host institution | Yes / No |
| Accommodation | Yes / No |
| Field visits | Yes / No |
| Materials used during the IP | Yes / No |
| Social programmes | Yes / No |
| Other (please specify): ………………………….. | |

| 7. Your personal experience – evaluation of the IP |
|---|

Judgement of academic/learning outcomes of the IP:
 Scale 1-5: 1=poor/negative, 5=excellent       1 – 2 – 3 – 4 – 5
Judgement of personal outcomes of the IP:
 Scale 1-5: 1=poor/negative, 5=excellent       1 – 2 – 3 – 4 – 5
Did you encounter any serious problems during the IP?
 Scale 1-5: 1=not at all, 5=very much        1 – 2 – 3 – 4 – 5
                  If yes, please specify:
                  ……………………….
How satisfied were you with the academic activities and the pedagogical aspects of the IP in terms of the following aspects?
 Scale 1-5: 1=not at all, 5=very much

| The number of hours taught | 1 – 2 – 3 – 4 – 5 |
|---|---|
| The equipment used | 1 – 2 – 3 – 4 – 5 |
| The capabilities and expertise of the professors | 1 – 2 – 3 – 4 – 5 |
| The overall quality of teaching | 1 – 2 – 3 – 4 – 5 |
| The expected learning outcomes | 1 – 2 – 3 – 4 – 5 |
| The activities besides the general course | 1 – 2 – 3 – 4 – 5 |

Do you think participation in the IP will help you in your further studies/career?
 Scale 1-5: 1=not at all, 5=very much        1 – 2 – 3 – 4 – 5
Do you think participation in the IP will help you in finding a job?
 Scale 1-5: 1=not at all, 5=very much        1 – 2 – 3 – 4 – 5
Overall evaluation of the IP:
 Scale 1-5: 1=poor/negative, 5=excellent       1 – 2 – 3 – 4 – 5
Recommendations and ideas for the IP organisers:
……………………………

**ACKNOWLEDGMENTS**

The authors would like to thank the companies *Desguace Malvarrosa*, *Elasis*, *Mann+Hummel GmbH*, *Dulcesol*, *Centrale del Latte di Salerno*, *Nicolas Feuillatte* and their staff for the help to develop SAVRO and HUMAIN Intensive Programs. An appreciation is also due to all the students that attended the six implementations of the intensive programmes.